\documentclass[review]{elsarticle}

\usepackage{lineno,hyperref}
\modulolinenumbers[5]

\journal{a journal}

\begin{document}

\begin{frontmatter}

\title{An Empirical Study on End-to-End Singing Voice Synthesis with Encoder-Decoder Architectures}



\author[addr1]{Dengfeng Ke}
\ead[url]{dengfeng.ke@blcu.edu.cn}

\author[addr2]{Yuxing Lu}
\ead[url]{yxlu@stu.suda.edu.cn}

\author[addr3]{Xudong Liu\corref{corr_author}}
\ead[url]{liuxudong.bjfu@outlook.com}
\cortext[corr_author]{Corresponding authors}

\author[addr3]{Yanyan Xu}
\ead[url]{xuyanyan@bjfu.edu.cn}

\author[addr4]{Jing Sun}
\ead[url]{jing.sun@auckland.ac.nz}

\author[addr4]{Cheng-Hao Cai\corref{corr_author}}
\ead[url]{chenghao.cai@auckland.ac.nz}

\address[addr1]{School of Information Science, Beijing Language and Culture University, China}
\address[addr2]{School of Computer Science and Technology, Soochow University, China}
\address[addr3]{School of Information Science and Technology, Beijing Forestry University, China}
\address[addr4]{School of Computer Science, the University of Auckland, New Zealand}

\begin{abstract}
With the rapid development of neural network architectures and speech processing models, singing voice synthesis with neural networks is becoming the cutting-edge technique of digital music production. In this work, in order to explore how to improve the quality and efficiency of singing voice synthesis, in this work, we use encoder-decoder neural models and a number of vocoders to achieve singing voice synthesis. We conduct experiments to demonstrate that the models can be trained using voice data with pitch information, lyrics and beat information, and the trained models can produce smooth, clear and natural singing voice that is close to real human voice. As the models work in the end-to-end manner, they allow users who are not domain experts to directly produce singing voice by arranging pitches, lyrics and beats.
\end{abstract}


\end{frontmatter}

\linenumbers

\section{Introduction}

Singing is a way for people to spread their thoughts and emotions, and it is similar to but not the same as speaking. Despite recent paradigms in speech synthesis systems based on machine learning, e.g., WaveNet \cite{DBLP:conf/ssw/OordDZSVGKSK16}, Tacotron \cite{DBLP:conf/interspeech/WangSSWWJYXCBLA17, DBLP:conf/icassp/ShenPWSJYCZWRSA18} and FastSpeech \cite{DBLP:conf/nips/RenRTQZZL19,DBLP:conf/iclr/0006H0QZZL21}, the application of machine learning to Singing Voice Synthesis (SVS) remains a challenging task. A typical speech synthesis system employs a front-end model to map text data to speech information, followed by a vocoder to convert the speech data into an audio format. Speech synthesis systems, however, cannot directly synthesize singing voices because they are influenced by several factors such as pitch, rhythm, emotion, and singing style. If we directly apply speech synthesis technology to SVS, the following problems may occur:
\begin{itemize}
\item It is difficult to match text to a song. Two adjacent phonemes are mutually constrained and difficult to distinguish because most phonemes of speech are independent of each other, whereas the phonemes of song obey the melodic continuity.
\item The phonemes' acoustic characteristics are more variable, influenced by the pitch, rhythm and strength of melody.
\item SVS uses more phonetic symbols than speech synthesis. Phonetic symbols in speech synthesis only contain vowels and consonants, as well as a few stress and tone changes. However, in SVS, phonetic symbols also need to include pitch information.
\end{itemize}

SVS has evolved into a hybrid of speech synthesis and electronic music. Moreover, multi-lingual engines such as VOCALOID \cite{DBLP:conf/interspeech/KenmochiO07}, and Sinsy \cite{DBLP:conf/apsipa/HonoMNHONT18} have been developed. However, voice synthesized by these engines usually sounds unnatural as four challenging problems need to be solved:
\begin{itemize}
\item How to map a musical score containing textual information to high-quality acoustic features?
\item How to decode the acoustic features into an audio digital signal format?
\item How to build efficient SVS workflow that does not rely on the collaboration of experts from different fields, including text grammar rules, music theory, digital signal processing and statistical machine learning?
\item How to reduce the accumulation of modular flaws that degrade the whole model's performance and cause poor synthesis quality?
\end{itemize}

To solve the above-mentioned problems, this paper presents an empirical study on end-to-end SVS models, especially how to reduce model development costs and improve the quality of synthesized audio. The contributions of this work are as follows:
\begin{itemize}
\item We implement an end-to-end SVS front-end model based on the FastSpeech \cite{DBLP:conf/nips/RenRTQZZL19}, HifiSinger \cite{DBLP:journals/corr/abs-2009-01776} and XiaoiceSing \cite{DBLP:conf/interspeech/LuWL0Z20} architectures. The model can reduce the complexity of building SVS systems and avoid error superposition and module mismatch between different modules.
\item In order to obtain better sound quality, we improve a number of vocoders according to the characteristics of SVS tasks.
\item We build a Chinese SVS system based on the front-end model and improved vocoders.
\item We compare and select acoustic features in the SVS system to match the front-end model and vocoder, which further improves the system performance.
\end{itemize}
The structure of this paper is as follows: Section 2 reviews sequence-to-sequence neural architectures. Section 3 presents the front-end SVS model. Section 4 introduces a variety of commonly used vocoders and our improvement methods for these vocoders. Section 5 presents experimental results on Chinese SVS tasks and tests the combination of the front-end model and improved vocoders. Section 6 gives the conclusion of our work.

\section{Sequence-to-Sequence Neural Architectures}

This section reviews sequence-to-sequence models used in this work, including Encoder-Decoder architectures, embedding methods, conditional mechanisms, attention mechanisms, Transformer and end-to-end speech synthesis. This section will focus on the relationship between the above techniques and Seq2Seq tasks, which refers to mapping input feature sequences to output feature sequences.

\subsection{Encoder-Decoder Models}

The difficulty of Seq2Seq tasks is the variable lengths of input and output sequences, making it difficult to learn the mappings from input sequences to output sequences using classical feed-forward neural networks. A common approach for solving the variable-length Seq2Seq problem is to use an Encoder-Decoder structure \cite{DBLP:conf/emnlp/ChoMGBBSB14} based on Recurrent Neural Networks (RNN) \cite{DBLP:journals/cogsci/Elman90}. The central idea of Encoder-Decoder is to encode input features into a fixed-length intermediate vector $e$, and then decode $e$ into output features. In this structure, the input and output can be serializable features, e.g., text, images, speech and tags. The objective of model parameter optimization is the numerical mapping relationship between the input features and the output features. Since the dimension of the intermediate vector e is fixed, the lengths of the input and output sequences can be different, eliminating the need for input and output sequence alignment.

The Encoder-Decoder architecture has been applied successfully to the Tacotron 2 speech synthesis system \cite{DBLP:conf/icassp/ShenPWSJYCZWRSA18}. The Encoder of Tacotron 2 encodes the input text as digital features using character embedding, smoothes the contextual information using a convolutional layer, and then converts it to intermediate variables using a bidirectional Long Short-Term Memory (LSTM) \cite{DBLP:journals/neco/HochreiterS97,DBLP:journals/tsp/SchusterP97} layer. The Decoder of Tacotron 2 decodes intermediate vectors into acoustic features by two LSTM layers, a linear layer and a post-net layer.

\subsection{Embedding Methods}

In this work, word embedding and position embedding are used to encode text and pitch information. When training neural networks, embedding methods can map an M-dimensional feature to another N-dimensional feature vector in the Euclidean space \cite{DBLP:conf/nips/MikolovSCCD13}. In Encoder-Decoder structures, sequences are usually encoded with an RNN along the time dimension. At each step of the computation, the RNN adds the previous information and passes the computation results forward. The RNN is slow because the loop-based training algorithm runs iteratively over variable-length sequences, increasing the computation and training time of the RNN. This problem can be solved by adding position information into input features and replacing the RNN with a feed-forward neural network.

Position encoding can have different forms. For instance, the fixed sinusoidal positional embedding can utilize the periodicity of the sinusoid to achieve relative positional encoding \cite{DBLP:conf/nips/VaswaniSPUJGKP17}. Since the length of the sequence is not a constraint for the sinusoidal function, it allows the model to learn the relative position and represent longer sequences than the training sequence.

\subsection{Condition Mechanisms}

A condition mechanism \cite{DBLP:conf/ssw/OordDZSVGKSK16} acts as a condition variable when a neural network model requires multiple data to act simultaneously on a single data stream. Assume there is a given input feature vector $\mathrm{X}=\left\{\mathrm{x}_{1}, \ldots, \mathrm{x}_{\mathrm{T}}\right\}$, the model's probability distribution can be treated as $\mathrm{P}(\mathrm{X})=\prod_{\mathrm{t}=1}^{\mathrm{T}} \mathrm{p}\left(\mathrm{x}_{\mathrm{t}} \mid \mathrm{x}_{1}, \ldots, \mathrm{x}_{\mathrm{t}-1}\right)$. When an additional input $h$ is given as the conditional control of the model, the probability distribution of the model is $\mathrm{P}(\mathrm{X} \mid \mathrm{h})=\prod_{\mathrm{t}=1}^{\mathrm{T}} \mathrm{p}\left(\mathrm{x}_{\mathrm{t}} \mid \mathrm{x}_{1}, \ldots, \mathrm{x}_{\mathrm{t}-1}, \mathrm{~h}\right)$, where $h$'s dimension is the same as $x_t$'s, and $h$ will act on every $x_t$ as the control condition of the whole model, which is customarily called the global condition. Given a condition variable $\mathrm{L}=\left\{\mathrm{l}_{1}, \ldots, \mathrm{l}_{\mathrm{t}}\right\}$, where $L$ has the same dimension as $X$, then at each moment, $x_t$ has a corresponding $l_t$ as the condition control, and $L$ is called the local condition. Adjusting the model's condition control can guide the model to generate predictive features of the specified sound style.

\subsection{Attention Mechanisms}

An attention mechanism \cite{DBLP:conf/nips/MnihHGK14} can give different weights to different features. It is usually added between Encoder and Decoder to solve the alignment problem between input and output sequences. For example, in machine translation tasks, the attention mechanism can solve the alignment problem of some words between different languages \cite{DBLP:journals/corr/BahdanauCB14}. The Multi-Head Attention (MHA) mechanism \cite{DBLP:conf/nips/VaswaniSPUJGKP17} is an extension of the attention mechanism by dividing attention into multiple subspaces and allowing the model to focus on the representation vectors in different subspaces at different positions.

\subsection{Transformer}

Transformer \cite{DBLP:conf/nips/VaswaniSPUJGKP17} was first used to solve the machine translation problem . It was later extended to a variety of Seq2Seq tasks, giving rise to the Bidirectional Encoder Representation from Transformers (BERT) \cite{DBLP:conf/naacl/DevlinCLT19} for natural language understanding, FastSpeech \cite{DBLP:conf/nips/RenRTQZZL19} for speech synthesis, etc. The Transformer model can be divided into two parts: Encoder and Decoder, where the input is encoded into digital features with position information via the Embedding layer in the Encoder. Then the output will be looped through a multi-layer network, including the MHA mechanism and a feed-forward network in each layer, with residual connections and regularization added after each layer of the network. The MHA mechanism is used to connect the Encoder and Decoder modules. The Decoder employs a cyclic autoregressive structure to splice the input with the output of each step as the input for the next iteration. After encoding the input information with positional encodings, the intermediate vector will be sent into a multi-layer loop that includes a combination of a masked MHA mechanism and a feed-forward network. Finally, the results pass through a linear layer and a Softmax layer to output the prediction probability.

\subsection{End-to-End Speech Synthesis}

In traditional machine learning systems, multiple independent modules are required to collaborate to solve a problem. Each module is trained individually to complete a separate task before being combined. The advantages of such systems are that they are more interpretable and can be trained in parallel. The disadvantages are that the accumulation of errors from multiple modules may increase the overall system error, and the design of different modules requires the involvement of professionals from various fields, as well as the collection and labeling of various training data. By contrast, in an end-to-end system, feature data are fed into one end of the system, and prediction results are obtained from the other end. The system only attempts to reduce the difference between the prediction results and the expected results, but is not concerned with the interpretability of the model. Therefore, the whole system can be regarded as a black box. The benefits of this approach are as follows. First, it avoids the collaborative work of multi-domain professionals and reduces the difficulty of model building. Second, it avoids the accumulation of errors among multiple modules. Last, it avoids the mismatch caused by different training data on different modules.

Tacotron \cite{DBLP:conf/interspeech/WangSSWWJYXCBLA17, DBLP:conf/icml/Skerry-RyanBXWS18} and Tacotron2 \cite{DBLP:conf/icassp/ShenPWSJYCZWRSA18} are end-to-end speech synthesis front-end models based on neural networks. Tacotron2 consists of an Encoder and a Decoder. The Encoder uses a Convolutional Neural Network (CNN) \cite{DBLP:journals/neco/LeCunBDHHHJ89} and a RNN to map indefinite-length text information into fixed-length intermediate vectors, where the CNN is used to smooth text features and the RNN is used to associate contextual feature vectors. The Decoder uses an autoregressive RNN to expand the intermediate vectors, leading to new vectors with acoustic feature lengths. After that, the new vectors are decoded into acoustic features using CNNs and fully connected neural networks. In the decoding process, an attention mechanism is added to determine the position of the acoustic features corresponding to each text feature. Attention mechanisms are crucial in speech synthesis models, and the choice of different attention mechanisms will lead to different training results. The backend model usually uses the WaveNet model \cite{DBLP:conf/ssw/OordDZSVGKSK16}, which is a baseline model commonly used in the speech synthesis research community.

\section{The SVS Front-End Model}

This section describes our implementation of the end-to-end SVS front-end model. The front-end model is adapted from the HifiSinger \cite{DBLP:journals/corr/abs-2009-01776}, XiaoiceSing \cite{DBLP:conf/interspeech/LuWL0Z20} and FastSpeech \cite{DBLP:conf/nips/RenRTQZZL19} models. The dataflow of the SVS model is as follows. Firstly, text features, including lyrics, pitch and beat, are extracted from a score. The lyrics information is divided into vowel and rhyme as the smallest unit of text according to the phonetic sound and rhyme rules, which is called Phone features. The Pitch information is annotated on each unit. Beat information is applied to calculate the duration of each unit, labeling the number of frames of pronunciation duration for each Phone, which constitutes the Duration feature. After that, the input features are predicted by the front-end model and the acoustic features of the singing voice are decoded by the vocoder. This method is based on the Chinese dataset and uses the vocal-rhyme approach to encode lyrics. Considering the future support for English and other languages, we use Phone feature to define the text minimum unit.

The front-end model is used to solve the Seq2Seq task of transforming the score information to the acoustic features of the singing voice. The model can be divided into three parts: Encoder, Decoder, and Length Regulator. The Encoder and Decoder with the MHA mechanisms are used to provide contextual association information, and the Length Regulator is used to provide alignment information. The Length Regulator uses beat information to calculate the mapping relationship between word and frame durations and connect two sequences of different lengths.

In the Encoder-Decoder Structure, the input is a sequence of Phones for the lyrics, which is first converted into vector features by the Embedding layer, after which the sinusoidal position encoding is added, and the local conditional Pitch information is added to the data stream after the Pitch condition control module. In practice, features are classified as global or local conditions according to their type and range of influence. The front-end model uses the Condition mechanism to guide the synthesis of acoustic features. To be specific, Pitch information corresponds to each Phone and acts on each character feature of a sample as a local condition. Speaker identity information corresponds to each Phone and acts as a global condition on a sample. The Encoder module is made up of a 6-layer structure, and each layer contains a self-MHA mechanism for extracting contextual association information and a CNN layer for smoothing features and de-linearization. The Decoder uses the same multi-layer MHA and CNN structure as the Encoder, and the output of the Decoder is the embedding vector of speech features. Additionally, the decoder has two optional modules. The first optional module is CBHG \cite{DBLP:conf/interspeech/WangSSWWJYXCBLA17}, which is used to convert Mel spectrograms \cite{mermelstein1976distance,IEEE1163420} to linear spectrograms. The second optional module is abbreviated as post-net \cite{DBLP:conf/icassp/ShenPWSJYCZWRSA18}, which is a post-processing network for further smoothing acoustic features.

The Length Regulator module is placed between the Encoder and the Decoder, which is used to expand the encoded text length features to audio length features by Duration information. It calculates the Duration information of each word in the lyrics based on the beat information, extracts the Duration information of each Phone, and learns the prediction model of a word-to-vocal duration simultaneously for the calculation of Duration information in the synthesized score. As the continuity of the singing voice between different Phones is stronger than that of the spoken voice, an additional smoothing operation is needed to make the final waveform more natural and effective. The smoothing operation uses a Gaussian distribution to initialize a convolutional kernel of size 3, which is applied to the Duration mask matrix after stitching and is used to smooth the frame connections between Phones. The convolutional kernel is applied to both the start and end of each frame.

The front-end model uses different objective functions in predicting different acoustic features. The Mean Squared Error (MSE) is used as the objective function for the prediction model of Mel spectrograms and Bark Frequency Cepstral
Coefficients (BFCC) \cite{shannon2003comparative}. For the prediction model of linear spectrogram, the weighted absolute loss is used as the objective function, with two weights for the priority frequency bands of the original samples and the predicted samples respectively. During the training process, the Adam algorithm \cite{DBLP:journals/corr/KingmaB14} with a dynamically decayed learning rate is used to optimize the model.

\section{Improvements on Vocoders}

The essential difference between SVS and speech synthesis is that the acoustic features focus on different frequency bands. Speech synthesis only needs to focus on the low-frequency part to ensure that the synthesized speech is clear and easy to understand, while SVS needs to ensure that the synthesized singing voice is clear, melodic and conforms to the given pitch. Thus, SVS requires a focus on the full frequency range at the same time. We ensure pitch accuracy by enhancing the pitch control in both the front-end model and the vocoder section. Section 3 has shown how the front-end model enhances pitch control. This section will discuss the effect of the vocoder on pitch control.

Vocoders can be classified into digital signal-based algorithms, e.g., World \cite{DBLP:journals/ieicet/MoriseYO16} and Griffin-Lim \cite{DBLP:conf/icassp/GriffinL83}, and neural network-based vocoder models, e.g., WaveNet \cite{DBLP:conf/ssw/OordDZSVGKSK16}, WaveGlow \cite{DBLP:conf/icassp/PrengerVC19} and LPCNet \cite{DBLP:conf/icassp/ValinS19}. We improve the standard WaveNet model by adding a pitch control module to enhance the Pitch information in the features. Moreover, we add the pitch control module before the gate activation unit of WaveNet and add the pitch condition information to original features and spectrograms. This model is called Pitch Condition WaveNet. It can convert the Phone level Pitch information to Frame level Pitch information according to the extended operation in the Length Regulator module. The Pitch and spectrogram features are used to train the same neural network with different hyper-parameters, and the neural network is used to learn the mapping relationships between the two features.

We designed two processes for independent vocoder synthesis and acoustic feature comparison. In the independent vocoder synthesis, we use acoustic features extracted from the original singing voice in the same dataset as training samples to train the vocoder alone, and the synthesis results are used to compare and analyze the ability of the individual vocoder to model the singing voice waveform data. After that, we combine the front-end model to train the vocoders for acoustic feature comparison. The parameters of the front-end model are fixed, the front-end predicted acoustic features are used as the training set to train the vocoder model, and the synthesis results are used to compare the expression ability of different acoustic features and the decoding ability of the vocoder. The neural network vocoders such as WaveGlow \cite{DBLP:conf/icassp/PrengerVC19}, LPCNet \cite{DBLP:conf/icassp/ValinS19} and \cite{DBLP:conf/ssw/OordDZSVGKSK16} are trained using the objective function suggested in their original papers, and all three vocoders use the Adam optimization algorithm \cite{DBLP:journals/corr/KingmaB14} with a fixed learning rate.

\section{Experiments}

This section provides experiments on Chinese SVS tasks. Section \ref{sec:exp_setting} introduces the dataset, experimental devices and evaluation criteria. Sections \ref{sec:exp_pitch_condition_module} and \ref{sec:exp_optional_modules} provide comparative experiments on the front-end model to demonstrate the model's performance with the pitch control module and the optional modules. Section \ref{sec:exp_vocoders} provides comparative experiments on both independent vocoder synthesis and the pitch control module to demonstrate the effectiveness of the improved vocoder. Section \ref{sec:exp_features} combines the front-end model and the vocoder for acoustic feature comparison experiments.

\subsection{Experimental Settings and Evaluation Criteria}
\label{sec:exp_setting}

We used an Intel Xeon(R) E5-2650 CPU combined with an Nvidia GTX 1080Ti GPU for training. The training dataset consists of Chinese songs with a total recording time of about 3 hours and about 1,000 sentences with segments of 3-8 seconds. The recording setting was a quiet indoor environment with a microphone. The audio sampling rate is 16kHz, the frame length is 50ms, and the frameshift is 10ms. The spectrogram uses the short-time Fourier transform, and the absolute value of it is a 513-dimensional linear spectrogram feature, which is passed through a number of filters to obtain the 80-dimensional Mel spectrograms, which is regularized between -4 and 4. The BFCC feature is extracted by forwarding the spectrogram obtained from the Fourier transform through the Bark filter to get the energy information of different frequency bands, converting the energy information to inverse spectrogram. Finally, DCT is used to de-correlate. In pitch feature extraction, 440Hz is denoted as the standard A1, and a pure octave is evenly divided into twelve semitones according to the twelve mean law, after which the pitch information in the score is mapped to an integer pitch sequence.

Experimental results are evaluated objectively based on parallel data and assessed subjectively based on parallel and non-parallel data in the same test environment by the same listener. The experimental comparison results are evaluated using standard criteria in the field of neural networks and speech such as mean square error and average opinion score. In some experiments, the audio is represented by speech spectrograms, where horizontal axis denotes time, and vertical axis denotes frequency. Additionally, we use pitch correctness as an additional objective evaluation criterion for the song synthesis system. The mean opinion scores are ground truth for the authentic song recordings, recons (Mel) and recons (Linear) for the song reconstructed by the Griffin-Lim algorithm \cite{DBLP:conf/icassp/GriffinL83} using the Mel spectrogram or linear spectrogram extracted from the authentic song recordings, which are used in the experiments to compare the quality of the song with the reconstructed phase information.

Mean Square Error (MSE) is an objective measure of the difference between prediction and truth, which can reflect the fitting ability of the model. Due to the high precision of MSE, we keep a 4-decimal place for MSE in the experimental results.

\begin{table}[htbp]
\centering
\caption{The Marking Sheet of Mean Option Score (MOS)}
\begin{tabular}{lll}
\hline
MOS & Voice Grade & Voice Quality                                                           \\
\hline
5         & Excellent   & Smooth and natural, no distortion                                       \\
4         & Good        & Distortion is not obvious and not easy to detect                        \\
3         & Fair        & Obvious distortion, acceptable natural and clarity  \\
2         & Poor        & Serious distortion, unnatural, tiring to hear                           \\
1         & Bad         & Very poor voice quality, unacceptable hearing                           \\
\hline
\end{tabular}
\label{tab:mos}
\end{table}

Mean Opinion Score (MOS) is used to score the overall satisfaction of speech, and its score table is shown in Table \ref{tab:mos}, which is one of the important criteria for the subjective evaluation of speech. Listeners are asked to rate the sample audio in the same test environment, and the mean value of the scores is taken as the MOS score. To avoid excessive variance fluctuations, the sample audio needs to be rich enough, including parallel and non-parallel speech. In digital speech signals, the MOS score of 4.0-5.0 is usually considered as high-quality digital speech, the MOS score of 3.0-4.0 is considered as the speech quality that can meet the communication requirements, and the MOS score below 3.0 is considered as the synthetic speech quality, which is intelligible but not natural. The MOS score in the experimental results is kept to 2 decimal places.

Pitch Accuracy (PA) is used to measure the pitch accuracy of the predicted voice objectively by comparing the pitch information of the original voice with that of the predicted voice. Pitches within the semitone range of the original speech are considered as accurate, and pitches outside this error range are considered as error points. The experiment results keep two decimal places when counting the mean PA of the test set.

\subsection{Experiments on the Pitch Condition Module}
\label{sec:exp_pitch_condition_module}

In order to control the pitch of the synthesized audio, our SVS model uses the pitch condition module to utilize the pitch information as a condition on the front-end model. This approach reduces the amount of data required for the front-end model and the complexity of model construction. The following experiment compares the effect of adding the pitch control module at different stages of the data stream on the synthesis results.

To measure to what extent the pitch condition module affects the front-end model, we set three independent experiments for comparison, with the model fixed except for the pitch condition module. The data randomization factor is fixed to ensure the consistency of the training data, and the Mel spectrogram is used for the acoustic features. The following strategy is used to adjust the pitch condition module:
\begin{itemize}
\item Add the pitch condition module separately before the encoder module, the pitch information and the encoder input are both phone sequence lengths, and the encoded pitch information is added directly to the input, called ``encoder-only" in the following passage.
\item Add the pitch condition module separately to the decoder module, flatten the pitch information to the frame time length by extending and smoothing the length regulator module. Then add it to the input of the decoder module, which is called ``decoder-only" in the following passage.
\item Add pitch condition module to encoder and decoder modules at the same time. Apply the pitch information as the input of encoder and decoder modules with the two strategies above, which are called ``both" in the following passage.
\end{itemize}
The objective of this experiment is to compare the prediction ability of front-end models for acoustic features. The objective evaluation criteria use the MSE and PA of Mel spectrogram features to measure the difference between predicted features and real features, and the subjective evaluation criteria use the MOS to measure the quality of predicted synthesized song. The model predicts the Mel spectrogram using the inverse Mel (IMel) filter \cite{paper:imel}. The Griffin-Lim algorithm \cite{DBLP:conf/icassp/GriffinL83} is used to synthesize the audio for comparing the model synthesis effect.

The experimental results are shown in Table \ref{tab:result2}. The encoder only model achieves the best results in both pitch accuracy and Mel-MSE. In term of the MOS score, the three models does not show obvious differences, and the results of both and encoder only models are very close. It can also be found in the table that the loss convergence of the three models differs significantly.

\begin{table}[htbp]
\centering
\caption{Experimental results of pitch condition module}
\begin{tabular}{llll}
\hline
Model                 & Mel-MSE         & PA (\%)      & MOS                                      \\
\hline
Decoder only          & 0.0894          & 86.63          & 2.45$\pm$0.01           \\
\textbf{Encoder only} & \textbf{0.0323} & \textbf{93.91} & 2.58$\pm$0.11          \\
Both                  & 0.0581          & 82.28          & \textbf{2.59$\pm$0.10}  \\
Recons (Mel)           & /               & /              & 2.67$\pm$0.11           \\
\hline
\end{tabular}
\label{tab:result2}
\end{table}

\begin{figure}[htbp]
\centering
\includegraphics[scale=0.8]{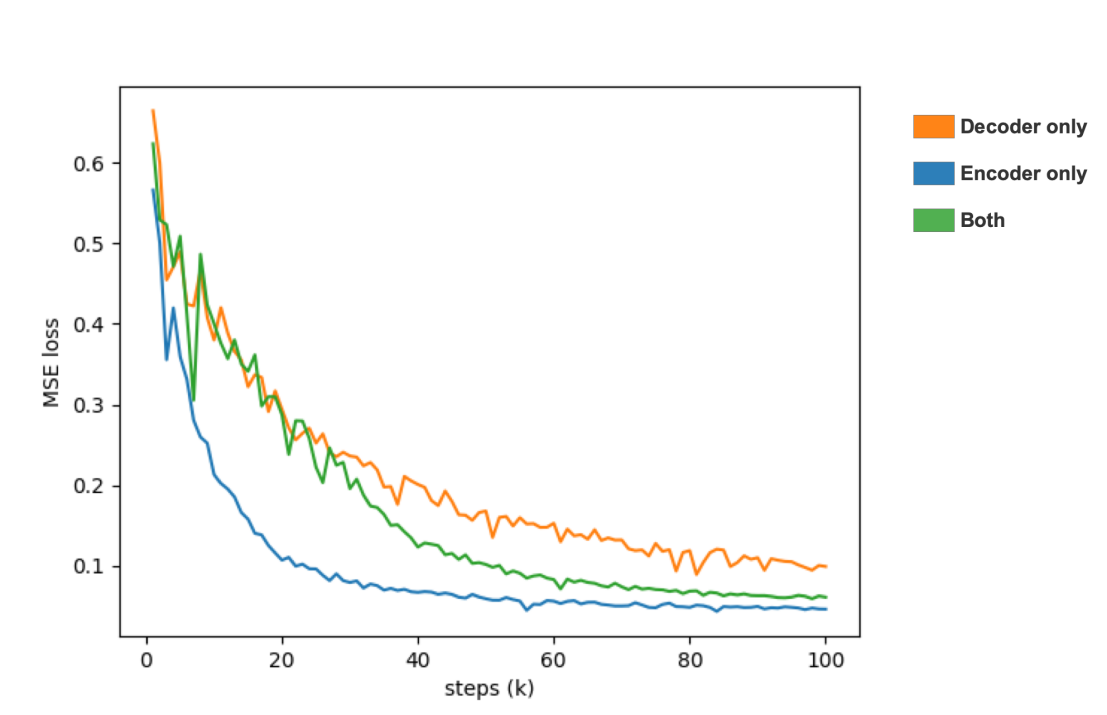}
\caption{Learning Curves of Pitch Condition Module}
\label{fig:3}
\end{figure}

Figure \ref{fig:3} shows the model loss curves. When the pitch condition module is applied to only the encoder, the model has a smooth loss curve finally reaches the lowest loss value. When the pitch condition module is applied to only the decoder, or both the encoder and the decoder, the curves have relatively frequent fluctuations. The experimental results show that applying the pitch condition module only to the encoder can improve the pitch quality of the synthesized song and speedup the training process.

\subsection{Experiments on the Optional Modules}
\label{sec:exp_optional_modules}

\subsubsection{Post-net}

\begin{figure}[htbp]
    \centering
    \includegraphics[scale=0.8]{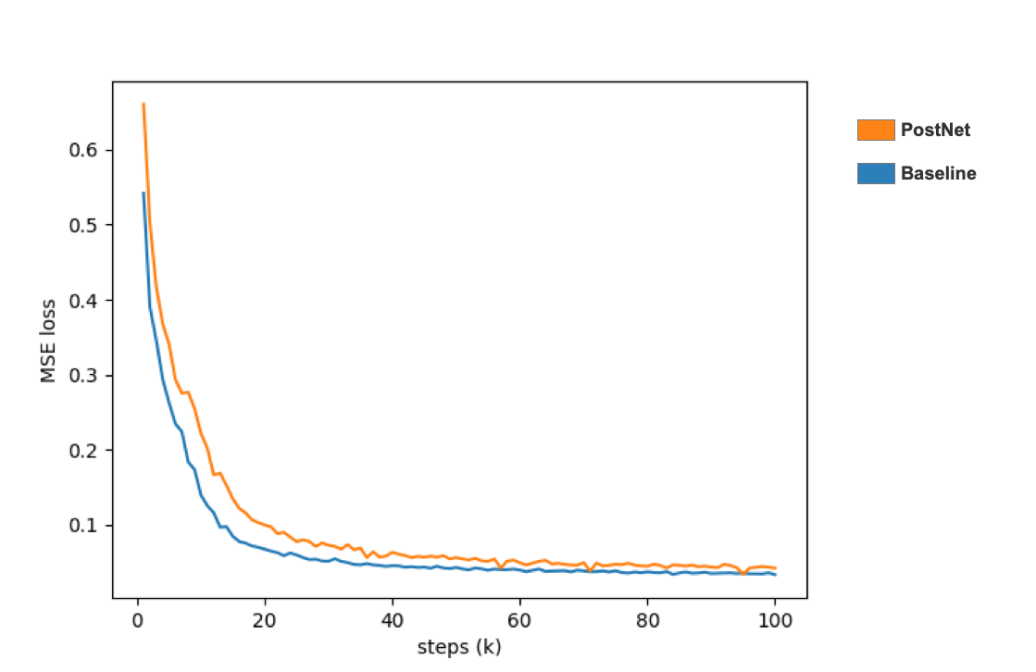}
    \caption{Learning Curves of the Post-net Module}
    \label{fig:4}
\end{figure}

We use the post-net of Tacotron 2 \cite{DBLP:conf/icassp/ShenPWSJYCZWRSA18} to smooth the synthesised audio. The encoder-only model in Section \ref{sec:exp_pitch_condition_module} is considered as the baseline model, and we train a front-end model with the post-net to compare with the baseline model. The objective evaluation criterion uses the MSE of the Mel spectrogram features to measure the difference between the predicted features and the true features, and the subjective evaluation criterion uses the speech spectrogram to observe the extraction of harmonics. Figure \ref{fig:4} shows the curves of model loss convergence. In the model with the post-net, the initial loss and loss convergence is not as good as the baseline model, indicating that the post-net does not improve the front-end model. This is probably because the front-end model already has the MHA mechanisms and convolution operators that can better smooth the synthesised features.

\subsection{CBHG}
\label{sec:CBHG}

The CBHG can achieve fast SVS by slightly losing sound quality, which is useful for low-resolution SVS tasks. In this section, we compare the CBHG with the IMel filter on the baseline encoder-only model in Section \ref{sec:exp_pitch_condition_module}. The objective evaluation criterion uses MSE of linear spectral features to measure the difference between the predicted features and the true features, and the subjective evaluation criterion uses MOS to measure the quality of the fast synthesized song and extracts the speech spectrogram to observe the synthesized harmonic effects.

\begin{figure}[htbp]
    \centering
    \includegraphics[scale=0.8]{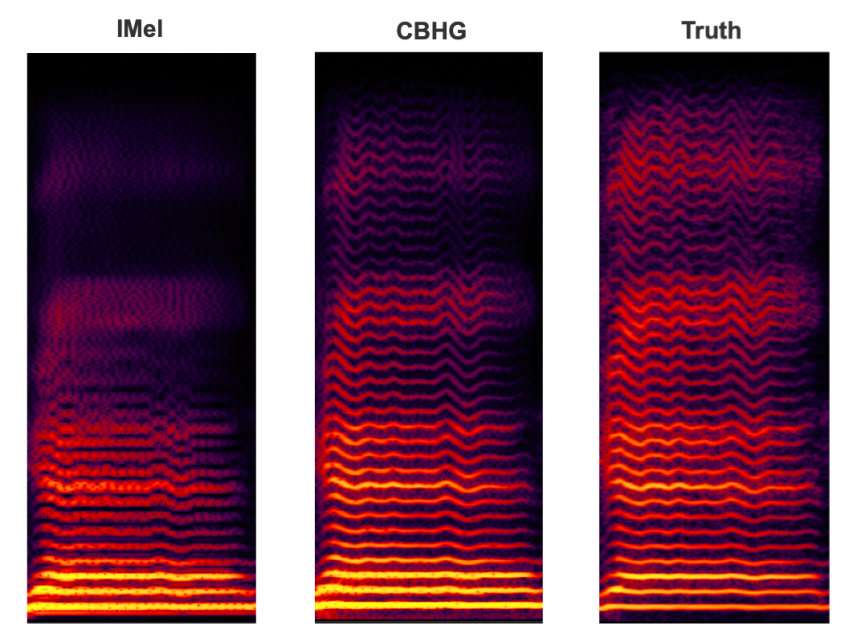}
    \caption{Spectrograms of CBHG Module}
    \label{fig:5}
\end{figure}

The experimental results in Figure \ref{fig:5} show that due to the characteristics of dense low-frequency and sparse high-frequency of the Mel filter, high-frequency information is lost when the linear spectrogram is recovered using the IMel filter, but this problem can be avoided by the CBHG. The MOS scores in Table \ref{tab:result4} show that the CBHG network layer can improve the synthesis quality of the linear spectrogram compared to the IMel filter. Moreover, the model with the CBHG can achieve a MOS score of 2.98, which is very close to the real speech.

\begin{table}[htbp]
\centering
\caption{MOS of CBHG module}
\begin{tabular}{ll}
\hline
Model          & MOS                 \\
\hline
Baseline+IMel  & 2.58$\pm$0.11           \\
\textbf{CBHG}  & \textbf{2.98$\pm$0.11}  \\
Recons (Linear) & 3.04$\pm$0.09           \\
\hline
\end{tabular}
\label{tab:result4}
\end{table}

\subsection{Experiments on Independent Vocoders}
\label{sec:exp_vocoders}

\begin{table}
\centering
\caption{Experimental Results of Independent Vocoder}
\begin{tabular}{llll}
\hline
Vocoder          & Training Time (D) & \# Points / Sec. & MOS                 \\
\hline
Griffin-Lim      & /                    & 87,600                                                                   & 3.04$\pm$0.09           \\
World            & /                    & \textbf{155,275}                                                         & 3.47$\pm$0.17           \\
WaveGlow         & 14                   & 13,4459 (GPU)                                                             & 2.96$\pm$0.11           \\
LPCNet           & \textbf{5}           & 40,483                                                                   & 3.34$\pm$0.12           \\
\textbf{WaveNet} & 10                   & 267 (GPU)                                                                & \textbf{3.82$\pm$0.12}  \\
GroundTruth      & /                    & /                                                                       & 3.97$\pm$0.09           \\
\hline
\end{tabular}
\label{tab:result5}
\end{table}

In this Section, we compare the effectiveness, training speed and synthesis speed of different vocoders. The objective evaluation metric uses the training time and synthesis speed to measure the engineering value of the model, and the subjective evaluation criterion uses the MOS to measure the naturalness of the recovered waveform. The experimental results are shown in Table \ref{tab:result5}. In terms of the training time, the LPCNet takes the shortest time, the WaveNet is medium, and the WaveGlow is the longest. The WaveGlow requires the longest training time because the inverse matrix is prone to gradient explosion during training. Besides, although the WaveNet is not superior in terms of synthesis speed due to its autoregressive characteristics, it can used to satisfy high-quality requirements. Additionally, the LPCNet can synthesize 16k audio at 2.5 times the real-time rate based on the CPU configuration.

In terms of audio quality, the World does not require training and achieves a relatively high score of 3.470. The audio synthesised by the Griffin-Lim has serious jitter and unnatural reverberation due to the need to estimate phase information, but the naturalness is higher and can decode lyrics and tones better. Moverover, the WaveNet achieves the best MOS score, which indicates that the autoregressive-based point prediction can synthesize any style of waveform under the guidance of the Mel spectrogram. Overall, the results show that among the neural network-based vocoders, the WaveNet is suitable for high-quality SVS applications, and the LPCNet is suitable for fast SVS applications.

\subsection{Experiments on the Pitch Condition WaveNet}
\label{sec:exp_wavenet}

In previous experiments on the front-end model, we have found that the pitch condition module is well suited for controlling pitch information in the SVS task. Moreover, the WaveNet model can decode acoustic features with high quality. In this section, we observe the effect of pitch condition model on the training speed and synthesis quality of WaveNet.

\begin{figure}[htbp]
    \centering
    \includegraphics[scale=0.7]{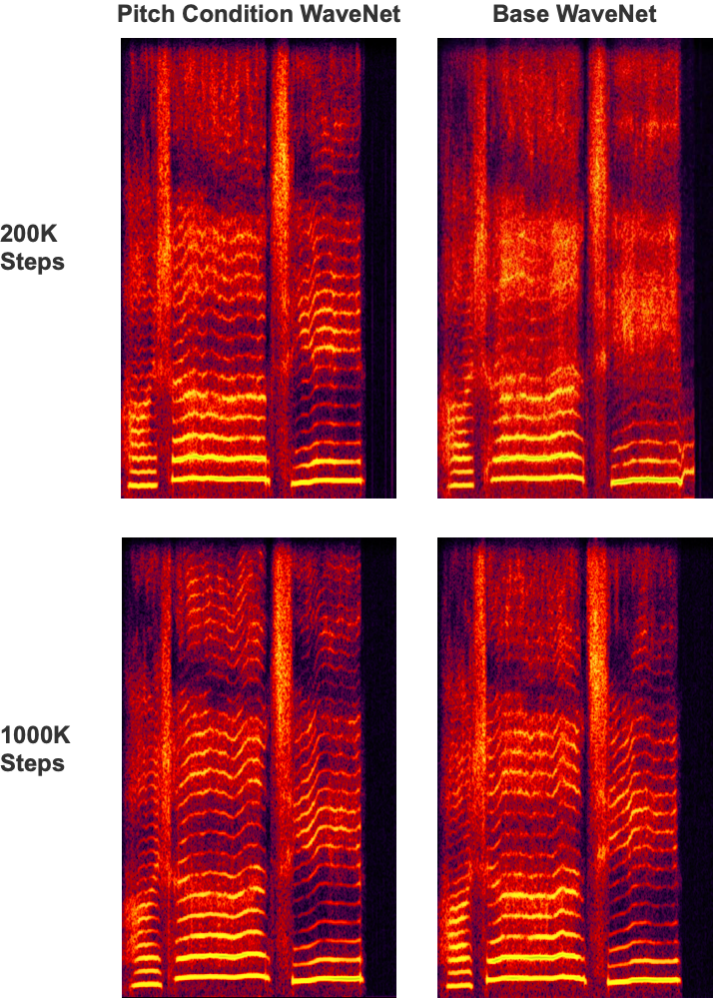}
    \caption{Spectrograms of Pitch Condition WaveNet}
    \label{fig:8}
\end{figure}

This experiment uses the WaveNet in Section \ref{sec:exp_vocoders} as the base WaveNet, and the modified pitch condition WaveNet is built by adding the local conditional pitch information into the activation gate of the base WaveNet. Both models are trained using the single-singer dataset. Figure \ref{fig:8} shows the spectrograms of waveforms reconstructed by the trained models. From 200k steps, the pitch condition WaveNet can resolve harmonics in the high-frequency part and the contours are close to the final result. However, for the base WaveNet, the high-frequency harmonics are unclear. At 1000k steps, the synthesis results of both models are close to each other, and some harmonic information is still lost in the high-frequency part of base WaveNet.

The experimental results show that the pitch condition WaveNet can resolve the waveform earlier than the base WaveNet, and the synthesis effect is more stable. Such improvement can accelerate the convergence speed of the model and enhance the stability. At the same time, this effect will gradually weaken with the increase of the training steps and finally achieve a comparable synthesis effect.

\subsection{Acoustic Feature Comparison}
\label{sec:exp_features}

In this section, we evaluate the quality of acoustic features, i.e., Mel spectrograms, linear spectrograms, [BCFF, Pitch] features and [F0, AP, SP] features, synthesised by the overall SVS system. The encoder-only model in Section \ref{sec:exp_pitch_condition_module} is used as the standard model and abbreviate as STD. We adapt the output function of the STD model to match the acoustic features, and the vocoders trained in Section \ref{sec:exp_vocoders} are used to decode the acoustic features. The PA is used to objectively evaluate the pitch accuracy of the synthesized vocals, and the MOS is used subjectively measure the naturalness of the recovered waveforms.

The experimental results are shown in Table \ref{tab:5}. The STD + WaveNet and STD + CBHG + Griffin-Lim systems achieve the highest pitch accuracy. Although Section \ref{sec:exp_vocoders} has demonstrated that the World vocoder can recover high-quality audio, the prediction of continuous F0 by the front-end model is still difficult, resulting in too smooth harmonics in the STD + World system with a pitch accuracy of only 77.56\%. In terms of the quality of sound, the STD + WaveNet system achieves a relatively high score of 3.57, which means that the STD + WaveNet system is suitable for high-quality SVS scenarios.

\begin{table}
\centering
\caption{Experimental Results of Acoustic Feature Synthesis}
\begin{tabular}{llll}
\hline
Model                        & Feature Type     & PA (\%)  & MOS                 \\
\hline
STD + World             & F0, AP, SP   & 77.56          & 2.65$\pm$0.13           \\
STD + WaveGlow          & Mel                   & 88.21          & 2.92$\pm$0.11           \\
STD + CBHG + Griffin-Lim & Linear                & \textbf{93.91} & 2.98$\pm$0.11           \\
STD + LPCNet            & BFCC, Pitch & 89.32          & 3.28$\pm$0.12           \\
\textbf{STD + WaveNet}  & \textbf{Mel}          & 93.33          & \textbf{3.57$\pm$0.13}  \\
Ground Truth                 & /                     & /              & 3.97$\pm$0.09           \\
\hline
\end{tabular}
\label{tab:5}
\end{table}

\section{Conclusion}

In this paper, we conducted a empirical study on end-to-end SVS models based on the encoder-decoder architectures. We implemented a SVS system that can use the alignment information, the lyrics and the beat information to synthesize audio. Based on the SVS system, we conducted experiments with different configurations of the front-end model and vocoders and obtained high-quality audio that is close to real human singing voice.

In the future, the SVS system may be further improved. First, although the CBHG network can improve the synthesis quality of the Linear spectrogram, it increases the computation time compared to the IMel filter. Considering the mapping relationship between the Linear spectrogram and the Mel spectrogram, it is possible to simply the model to a CNN or other simple models that better match the requirement of fast SVS. Second, the LPCNet model does not achieve the expected synthesis speed and sound quality possibly because the front-end model does not predict sufficiently precise acoustic features. We plan to make improvements to the front-end model to better support this vocoder. Finally, the WaveNet model achieves the best MOS in the experiments, but it is still time-consuming. We plan to speedup WaveNet using parallel computing techniques.

\section*{References}

\bibliography{mybibfile}

\end{document}